\documentclass[conference]{IEEEtran}
\IEEEoverridecommandlockouts
\usepackage{cite}
\usepackage{amsmath,amssymb,amsfonts}
\usepackage{algorithmic}
\usepackage{graphicx}
\usepackage{textcomp}
\usepackage{xcolor}
\usepackage[bottom]{footmisc}
\def\BibTeX{{\rm B\kern-.05em{\sc i\kern-.025em b}\kern-.08em
    T\kern-.1667em\lower.7ex\hbox{E}\kern-.125emX}}

\usepackage{booktabs}
\definecolor{ash}{RGB}{225, 225, 225}
\usepackage[hyphens]{url}
\usepackage{hyperref}

\begin{document}

\title{Can LLMs Generate Architectural Design Decisions? - An Exploratory Empirical study\\}

\author{\IEEEauthorblockN{Rudra Dhar}
\IEEEauthorblockA{\textit{Software Engineering Research Centre} \\
\textit{IIIT Hyderabad, India}\\
rudra.dhar@research.iiit.ac.in}
\and
\IEEEauthorblockN{Karthik Vaidhyanathan}
\IEEEauthorblockA{\textit{Software Engineering Research Centre} \\
\textit{IIIT Hyderabad, India}\\
karthik.vaidhyanathan@iiit.ac.in}
\and
\IEEEauthorblockN{Vasudeva Varma}
\IEEEauthorblockA{\textit{Language Technologies Research Centre} \\
\textit{IIIT Hyderabad, India}\\
vv@iiit.ac.in}
}
\maketitle

\begin{abstract}
Architectural Knowledge Management (AKM) involves the organized handling of information related to architectural decisions and design within a project or organization. An essential artefact of AKM is the Architecture Decision Records (ADR), which documents key design decisions. ADRs are documents that capture decision context, decision made and various aspects related to a design decision, thereby promoting transparency, collaboration, and understanding. Despite their benefits, ADR adoption in software development has been slow due to challenges like time constraints and inconsistent uptake. Recent advancements in Large Language Models (LLMs) may help bridge this adoption gap by facilitating ADR generation. However, the effectiveness of LLM for ADR generation or understanding is something that has not been explored. To this end, in this work, we perform an exploratory study which aims to investigate the feasibility of using LLM for the generation of ADRs given the decision context. In our exploratory study, we utilize GPT and T5-based models with 0-shot, few-shot, and fine-tuning approaches to generate the Decision of an ADR given its Context. Our results indicate that in a 0-shot setting, state-of-the-art models such as GPT-4 generate relevant and accurate Design Decisions, although they fall short of human-level performance. Additionally, we observe that more cost-effective models like GPT-3.5 can achieve similar outcomes in a few-shot setting, and smaller models such as Flan-T5 can yield comparable results after fine-tuning. To conclude, this exploratory study suggests that LLM can generate Design Decisions, but further research is required to attain human-level generation and establish standardized widespread adoption.
\end{abstract}


\section{Introduction}\label{sec:introduction}
\textit{Architectural Knowledge Management (AKM)} is the organized management of information on architectural decisions and designs within a project or organization. This involves capturing architectural knowledge, styles, design patterns, and quality attributes in a centralized repository. The focus is on ensuring traceability of decisions, promoting collaboration, facilitating knowledge reuse, and offering decision support. AKM aims to enhance communication, learning, and decision-making, contributing to the success of software development projects or other endeavours with complex architectures.

For several decades, the significance of AKM has been widely recognized, and various tools have been developed to assist AKM. \cite{b13}
Rainer et al. \cite{b15} state that while a huge amount of work is done to support AKM activities and capture Architectural Knowledge, it has not been sufficient. This has been a crucial reason restricting a wider adoption of AKM approaches, and more research is needed for automatically capturing this knowledge \cite{b14}.

An \textit{Architecture Decision Record (ADR)} is a crucial part of AKM. It entails the idea that software architecture is considered a set of Design Decisions \cite{b12}. It is a document used in software development to capture and document important \textit{Architecture Design
Decisions (ADD)}, made during the design and development of a software system. A detail explanation is given in Section \ref{sec:background}
Despite the well-established benefits of ADRs, their adoption has been slow to non-existent as described by Georg \textit{et al.} \cite{b27}. Unsuccessful adoption of ADRs in software development can occur due to several factors, including inadequate tool support, effort needed to capture Architecture Knowledge (AK), interruptions to the design process caused by documenting AK, and uncertainty regarding which AK needs documentation. \cite{b27}.

\textit{Large Language models (LLMs)} excel in comprehending contexts and generating text accordingly. Over the recent years due to advancement of LLMs, text generation has become more accessible. This paper delves into the exploration of whether LLMs can effectively generate Architectural Decision Records (ADRs). While the prospect of generating entire ADRs from a codebase remains a task for future endeavours, the focus of this work is on utilizing LLMs to generate Design Decisions from decisions Contexts as these are recognized as the core components of an ADR\footnote{https://docs.aws.amazon.com/prescriptive-guidance/latest/architectural-decision-records/adr-process.html}.

In the realm of \textit{Natural Language Processing (NLP)}, text generation is typically addressed through three distinct methodologies when utilizing LLMs. These approaches include \textit{zero-shot prompting, few-shot prompting}, and \textit{fine-tuning}. Section \ref{sec:background} provides a comprehensive exploration of the advantages and disadvantages associated with each approach \cite{b29}. It is recommended to initially employ zero-shot prompting, followed by few-shot prompting, and finally, fine-tuning due to the escalating complexity of implementation.

In this exploratory empirical study, we commence by defining the experimental subject, which involves gathering ADRs and choosing the LLMs for the study.
Next we perform experiments deploying the 3 approaches mentioned above.
In the zero-shot scenario, the model generates Decisions solely based on Context.
Whereas in the few-shot approach, the model is \textit{trained in-context} on a set of ADR examples.
Lastly, the paper delves into fine-tuning, where a generative model is \textit{trained} to produce Decisions from Contexts.

Our results show that LLMs do exhibit noteworthy capabilities in generating Architectural Design Decisions. While GPT-4 excels in 0-shot prompting, smaller models like text-davinci-003 and Flan-T5-base achieve comparable results with few-shot prompts and fine-tuning respectively. Smaller, fine-tuned models, like Flan-T5-base, requiring minimal hosting infrastructure, can prove useful for ADR generation within organizational infrastructure, particularly in privacy or security-sensitive scenarios.
We conclude that despite LLMs not being entirely dependable for ADR generation, they can effectively assist architects in documenting ADD.

Our experimental scripts and results are made available with an open-source license, to enable the independent verification and replication of the results presented in this study:\\
\url{https://github.com/sa4s-serc/ArchAI_ADR}

The remainder of this paper is structured as follows. Section II presents background concepts on ADR, LLM, and text generation. Section III presents the related work on tools and technologies for AKM and ADR. Section IV details the overall approach and the study design. Section V describes the results of the different experiments performed, and Section VI presents the related discussion. The threats to the validity of this study are thoroughly analyzed in Section VII. Finally, Section VIII documents our conclusions and future work.

\section{Background}\label{sec:background}
This section describes some key concepts and ideas that are used in this study, namely, ADRs and LLMs, and approaches of text generation.

\subsection{Architectural Decision Record (ADR)}
\noindent Software Architecture can be represented as a set of \textit{Design Decisions} \cite{b12}. An \textit{Architecture Decision Record (ADR)} is a software development document that captures and documents crucial architectural decisions made throughout a project. It encompasses information about the \textit{context} of the problem, the \textit{decision} taken, the anticipated consequences of the decision, and relevant references. ADRs are instrumental in promoting transparency, fostering collaboration, and preserving the historical background of architectural choices. They play a vital role in project documentation and ensuring that decision-making is well-informed. The primary elements of an ADR comprise the problem context and the associated decision.

\subsection{Large Language Models (LLM)}
A \textit{Language Model} is a probabilistic model of a natural language that can generate probabilities of a series of words. Whereas \textit{Large Language Models (LLM)} are similar probabilistic AI models trained on extensive data for understanding and generating text. They have billions of parameters and long Context Lengths, and are used for various NLP tasks. The Context Length of an LLM is the number of tokens it considers when processing or generating text. Tokens, in turn, are units of text, which might correspond to a word or a sub word. For example, in a sentence like "ChatGPT is great," the tokens are "Chat", "GPT," "is," and "great." Transformers, featuring attention mechanisms are the underlying architecture of these LLMs \cite{b8}. Transformers comprise two main components: an Encoder and a Decoder. The Encoder is responsible for handling input text, while the Decoder generates new text based on the encoded information. LLMs can take different forms: they can be Encoder-only models, exemplified by BERT \cite{b5}, feature an Encoder-Decoder architecture like T5 \cite{b10}, or be Decoder-only models, such as GPT \cite{b11}. GPT functions as a decoder only model in text completion, embodying an auto-regressive or causal LLM. In contrast, T5 is an encoder-decoder model, incorporating both an encoder for processing input text and a decoder for output text generation. In recent years, LLM's like GPT and T5 has revolutionised the world by making Natural Language Generation task easy, robust, and accessible to anyone. Throughout this paper, we use the terms GPT-chat for the models GPT-3.5-turbo and GPT-4, distinguishing them from other GPT models, which are referred to as text-completion models.

\subsection{Zero-shot, Few-shot and Fine-tuning approaches for text generation}

\textit{Zero-shot learning} involves training a model to perform a task without prior exposure to examples of that task. Few-shot learning introduced by Tom \textit{et al}. \cite{b28} extends this by allowing the model to be \textit{trained in-context} on a small number of examples for a specific task. \textit{Fine-tuning} entails further training of a pre-existing model on a particular task to enhance its performance on that task. Marius \textit{et al}. \cite{b29} studies these approaches in detail. Some important points, as described below should be considered while choosing the right approach.

Zero-shot and few-shot procedures, as defined, do not require training, in contrast to fine-tuning which explicitly demands it. The training process is expensive as it involves substantial investments in both hardware infrastructure and expertise. Additionally, zero-shot prompting operates without dependence on task-specific data, while few-shot prompting employs a modest number of task-specific examples. On the other hand, fine-tuning necessitates a considerable volume of labeled task-specific data for the training process. As a result, zero-shot and few-shot approaches demonstrate practical advantages, and the recommendation is to consider fine-tuning only if these approaches prove ineffective. Theoretically, a few-shot approach is supposed to generate better results than 0-shot prompt, whereas, fine-tuned models are supposed to generate even better response in the domain it's fine-tuned.

Nevertheless, the high-performing models designed for zero-shot and few-shot scenarios are typically huge generic LLMs boasting billions to trillions of parameters. Hosting such models locally can be challenging and, in many cases, impossible, leading to their reliance on cloud infrastructure accessible over the internet. This reliance on external services can give rise to privacy and legal concerns, particularly when private or proprietary data needs to be transmitted to third-party service providers. Conversely, smaller models tailored to specific tasks can achieve comparable results without the need for extensive resources. These more compact fine-tuned models can be deployed on a company's servers, mitigating some of the privacy and legal considerations associated with utilizing cloud services.
Our study aims to explore all these approaches with the different kind of generative LLM's mentioned above.

\section{Related work}\label{sec:related_work}
For several decades, the significance of Architecture Knowledge Management (AKM) has been widely recognized, and various tools have been developed to assist AKM. \cite{b13}
The focus on capturing and documenting Architecture Design Decisions (ADD) has grown with the recognition of ADDs' significance \cite{b23} \cite{b24}. Notable efforts, like Arman \textit{et al}.'s work on retrieving ADDs through the analysis of project history artifacts, have been made \cite{b25}.
However, these tools were largely deterministic and labour intensive and did not capture much attention from architects. However some Machine Learning based tools have began to emerged in the recent years.

Machine learning for Software Engineering has been a prominent topic in the past few years \cite{b16} \cite{b17}.
Generative AI is gaining popularity recently, particularly in tasks like automatic code documentation and summarization\footnote{https://paperswithcode.com/task/code-summarization}. While April \textit{et al.} worked on Code Documentation in Computational Notebooks \cite{b18}, Jian \textit{et al.} assembels multiples Foundational models for code summarization\cite{b21}.
Niche works, such as Intent-Driven Comment Generation by Fangwen \textit{et al}., have explored specific aspects of this field \cite{b19}.

Limited research exists in the realm of ML for Software Architecture or ADD. For instance, Xueying \textit{et al}. explored the Identification of Design Decisions from Mailing List using ML and NLP techniques predating the era of LLMs\cite{b20}.
In a similar vein, Manoj \textit{et al}. focused on extracting Design Decisions from natural language documentation, like Jira tickets \cite{b26}. It's noteworthy, however, that neither work did not specifically address ADRs and did not involve the generation of Design Decisions.

Despite these advancements, the field of Software Architecture remains relatively untouched by Generative AI or LLMs. In the context of ADR, Machine Learning applications have been limited, and to the best of our knowledge no works generating Design Decisions using LLMs exists.

\section{Study Design and Execution}\label{sec:study_design}
\begin{figure*}[t]
    \centering
    \includegraphics[width=\textwidth]{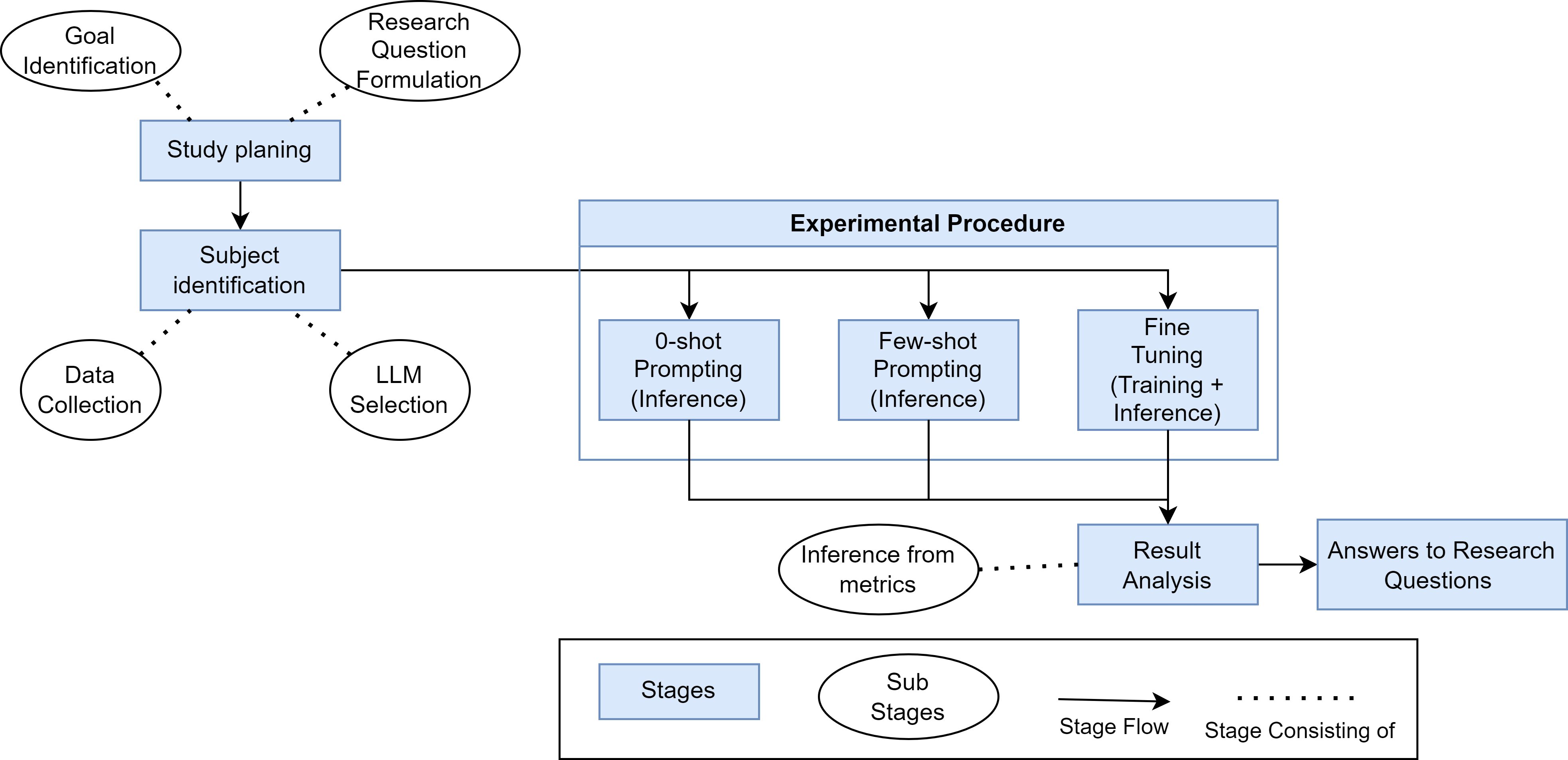}
    \caption{Study Design}
    \label{fig:study_diagram}
\end{figure*}

In this section, we document the empirical experiment executed for this study in terms of goal (Section IV-A), research questions (Section IV-B), study subject (Section IV-C), experimental procedure (Section IV-D), and experimental Metrics (Section IV-E). A visualization of the process is provided in Figure \ref{fig:study_diagram}.

\subsection{Goal}
This research is about an exploratory investigation to determine the feasibility of effectively utilizing LLMs for the generation of Architectural Design Decisions based on a provided context. The study aims to identify both the advantages and disadvantages associated with this approach.
More formally, by utilizing the Goal-Question-Metric approach \cite{b6}, this objective can be described as follows: \\
\textbf{Analyze} the effectiveness of Large Language Models\\
\textbf{For the purpose of} generating Architectural Design Decisions\\
\textbf{With respect to} contexts of Architecture Decision Records\\
\textbf{From the viewpoint of} Software Architects\\
\textbf{In the context of} using Generative AI for Architectural Knowledge Management

\subsection{Research Question}
In order to achieve our goal, we address the following three
research questions (RQ):

$\mathbf{RQ_1}$ \textit{Can LLMs be successfully employed to generate architectural design decisions from a given context in a zero-shot setting?}

Answering this introductory research question will help us determine if an architect can directly utilize a generic, foundational LLM, simply by presenting it with a decision Context, to obtain the corresponding Design Decision.

$\mathbf{RQ_2}$ \textit{Does few-shot approach affect or improve a LLM's ability to generate Design Decisions?} 

With $RQ_2$ we try to determine if LLMs can learn ADRs, \textit{in-context} and subsequently generate improved Design Decisions, by incorporating sample Context-Decision pairs into the prompt.
We investigate whether smaller and more cost-effective LLMs can effectively generate design decisions in few-approach at par more powerful models in 0-shot setting.

$\mathbf{RQ_3}$ \textit{Does Fine-tuning LLMs enhances it's capability of generating architectural Design Decisions based on a provided context?}

While research question $RQ_1$ and $RQ_2$ focuses on using foundational LLMs off the shelf, in $RQ_3$ we investigate if fine-tuning an LLM with Context-Decision pairs improves it's capability of generating Design Decision from Context. 
Huge LLMs capable of generating excellent text in zero and few-shot are mostly available as cloud service and cannot be trained or hosted loaclly. In $RQ_3$ we investigate if smaller models that can be trained and hosted locally can generate Design Decisions at par the extensive models after fine-tuning.

\subsection{Experimental Subject}

\subsubsection{ADR Data}
We performed an web-search with key-words \textit{Architecture Decision Record} and went through the top 20 results.
Among these we found set of repositories, mostly from GitHub. After reviewing, we chose five repositories with a substantial number of sample ADRs that adhered mostly to a standard format~\footnote{https://adr.github.io/madr/}. 
We got 17 ADRs from archane-framework\footnote{https://github.com/arachne-framework/architecture},
17 from winery\footnote{https://github.com/eclipse/winery/tree/d84b9-d7b6c9828fd20bc6b1e2fcc0cf3653c3d43/docs/adr},
32 from joelparkerhenderson\footnote{https://github.com/joelparkerhenderson/architecture-decision-record/tree/main/examples} repository,
14 from cardano\footnote{https://plutus-apps.readthedocs.io/en/latest/adr/}, and
15 from island\footnote{https://docs.devland.is/technical-overview/adr}.
We gathered a total of 95 ADRs from these sources
or GitHub repositories, we were able to download it directly. However, for other sources, we had to perform web crawling on the master repository to retrieve the corresponding ADRs.
To facilitate our experiments, we focused on extracting the Context and Decision components. This extraction process involved a combination of regular expressions and manual intervention. A sample ADR data after extracting Context-Decision is given in Figure \ref{fig:Context_Decision}.

\begin{figure}[t]
\centering
\fbox{
    \begin{minipage}{0.45\textwidth}
        \textbf{Context}\\
        We need to decide on whether to use Python as a programming language for our project. Our project involves data analysis, machine learning, and web development.
        \\
        
        \textbf{Decision}\\
        We have decided to use Python as our primary programming language for our project.
    \end{minipage}
}
\caption{Sample ADR after Extracting Context-Decision}
\label{fig:Context_Decision}
\end{figure}

\subsubsection{LLM}
\begin{table}[ht]
\centering
\caption{LLMs used in this study}
\begin{tabular}{lcccc}
    \toprule
    \textbf{family} & \textbf{model} & \textbf{size} & \textbf{context length} & \textbf{availability}\\
    \midrule
    & GPT2 & 124M & &\\
    GPT-2 & GPT2-medium & 335M & 1024 & local\\
    & GPT2-large & 774M & &\\
    & GPT2-xl & 1.5B & &\\
    \midrule
    GPT-3 & ada & 350M & 2048 &api\\
    & davinci & 175B & &\\
    \midrule
    GPT-3.5 & text-davinci-003 & 175B & 4000 & api\\
    & GPT-3.5-turbo & 175B & &\\
    \midrule
    GPT-4 & GPT-4 & T+ & 8192 &api\\
    \midrule
    & T5-small & 60M & &\\
    T5 & T5-base & 223M & infinite & local\\
    & T5-large & 738M & &\\
    & T5-3b & 3B & &\\
    \midrule
    T0 & T0-3b & 3B & infinite & local\\
    \midrule
    & Flan-T5-small & 77M & &\\
    Flan-T5 & Flan-T5-base & 248M & infinite & local\\
    & Flan-T5-large & 783M & &\\
    & Flan-T5-xl & 3B & &\\
    \bottomrule
\end{tabular}
\label{tab:LLM}
\end{table}

Numerous LLMs are available from various organizations and we tried to get a comprehensive representation of them. Generative LLMs can be of two categories: encoder-decoder, and decoder only. In this study, we opted for two widely recognized model series: GPT by OpenAI, representing Decoder-only models, and T5 by Google, representing Encoder-Decoder models. Within these series, we have incorporated models of varying sizes, training data, training techniques and interaction styles. For instance, we varied the model sizes from small to extra large, for T5 and Flan-T5 models. GPT-3 and GPT-3.5 differ in both training data and methods. Additionally, GPT offers text-completion and chat models, both included in this study. 

Table~\ref{tab:LLM} presents the essential characteristics of the models employed in this study. "Size" and "Context Length"  are explained in Section~\ref{sec:background}. The "Availability" category indicates whether the model can be hosted on a local PC or server, or if an API call is required to access the model from the service provider.

It is crucial to acknowledge that ADRs are extensive texts, often comprising thousands of characters. However, transformer-based models like GPTs have constraints on context length. Consequently, we selectively utilized data that falls below the context length for each model (refer Table~\ref{tab:LLM}), ensuring compatibility. Notably, T5-based models do not face this limitation, as T5 can handle infinite context lengths owing to its relative positional encoding. Nevertheless, owing to hardware constraints, we found it necessary to establish a threshold even for T5 models. Additionally, during fine tuning, there was a necessity to exclude data beyond a specific threshold due to hardware constraints.

\subsection{Experimental Procedure}
\subsubsection{0-shot approach}
Here we provided the model with the decision \textit{Context} along with a relevant prompt, anticipating it to produce the desired \textit{Design Decision}. 
We experimented with several prompt variations on a subset of samples. For instance, in one prompt, we solely provided the \textit{Context} without additional information, while in another, we included the phrase \textit{Architectural Decision Record} within the prompt. 
Through manual experimentation with various prompts on a subset of samples, we identified the most effective prompts which consistently generated correct and coherent Design Decisions, in proper format. Some noteworthy observations made in this phase are given in Section \ref{sec:discussion}. For GPT text-completion models, and T5 based models the most effective prompt is given in Figure \ref{fig:zero_shot_text-completion} and for GPT-chat models, and the most effective prompt is given in Figure \ref{fig:zero_shot_chat}.
\begin{figure}[ht]
\centering
\fbox{
    \begin{minipage}{0.4\textwidth}
        Architectural Decision Record \\
        \#\# Context: \\
        \{context\} \\
        \#\# Decision:
    \end{minipage}
}
\caption{0-shot prompt for GPT text-completion models, and T5 based models}
\label{fig:zero_shot_text-completion}
\end{figure}

\begin{figure}[ht]
\centering
\fbox{
    \begin{minipage}{0.4\textwidth}
        \{
        "role": "system", \\
        "content": "This is an Architectural Decision Record for a software. Give a \#\# Decision corresponding to the \#\# Context provided by the User "
        \} \\
        \{
        "role": "user", \\
        "content": \{context\}
        \}
    \end{minipage}
}
\caption{0-shot prompt for GPT-chat models}
\label{fig:zero_shot_chat}
\end{figure}

\subsubsection{few-shot approach}
In this approach, we engaged in \textit{in-context learning} by manually selecting two ADRs as gold samples. These gold samples were selected by one of the authors who played the role of an expert Software Architect. These gold samples, along with the context of the desired sample, were provided to the model. The expectation was for the model to learn in-context from the gold samples and predict the desired Decision based on the Context from the given sample.

Here also we performed manual experimentation with various prompts on a subset of samples and identified the most effective prompts which consistently generated correct and coherent Design Decisions, in proper format. For GPT text-completion models, and T5 based models the most effective prompt is given in Figure \ref{fig:few_shot_text-completion} and for GPT-chat models, and the most effective prompt is given in Figure \ref{fig:few_shot_chat}.
\begin{figure}[h]
\centering
\fbox{
    \begin{minipage}{0.4\textwidth}
        \#\# Context: \\
        \{gold context 1\} \\
        \#\# Decision: \\
        \{gold decision 1\} \\
        \#\# Context: \\
        \{gold context 2\} \\
        \#\# Decision: \\
        \{gold decision 2\} \\
        \#\# Context: \\
        \{context\} \\
        \#\# Decision:
    \end{minipage}
}
\caption{few-shot prompt for GPT text-completion models, and T5 based models}
\label{fig:few_shot_text-completion}
\end{figure}

\begin{figure}[h]
\centering
\fbox{
    \begin{minipage}{0.4\textwidth}
        \{
        "role": "system", \\
        "content": "These are architecture decision records. Follow the examples to get return Decision based on Context provided by the User"
        \} \\
        \{
        "role": "user", \\
        "content": "\#\# Context \{gold context 1\}"
        \} \\
        \{
        "role": "assistant", \\
        "content": "\#\# Decision \{gold decision 1\}"
        \} \\
        \{
        "role": "user", \\
        "content": "\#\# Context \{gold context 2\}"
        \} \\
        \{
        "role": "assistant", \\
        "content": "\#\# Decision \{gold decision 2\}"
        \} \\
        \{
        "role": "user", \\
        "content": "\#\# Context \{context\}"
    \end{minipage}
}
\caption{few-shot prompt for GPT-chat models}
\label{fig:few_shot_chat}
\end{figure}

\subsubsection{fine-tuning approach}
\begin{figure*}[t]
    \centering
    \includegraphics[width=\textwidth]{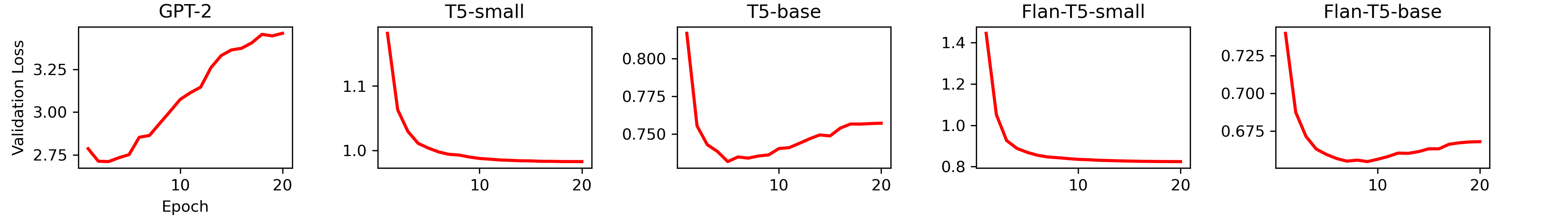}
    \caption{Training: validation loss / epoch}
    \label{fig:validation_loss}
\end{figure*}

For fine-tuning we used GPT2 and four T5 based models. GPT2 is a text completion model. Hence for GPT2, we concatenated each context with the corresponding decision and trained the model on generating the full text. However T5 is an encoder-decoder model. Hence the Context was provided to the Encoder, and the Decoder was supposed to generate the Decision. The model was trained on generating the correct Decision. A visual representation is given in Figure \ref{fig:Training_Format}
\begin{figure}[h]
\centering
\fbox{
    \begin{minipage}{0.4\textwidth}
        \textbf{GPT2} \\
        Generate: \{context\} \{decision\}\\
        \\
        \textbf{T5} \\
        Read: \{context\} \\
        Generate: \{decision\}
    \end{minipage}
}
\caption{Training Format}
\label{fig:Training_Format}
\end{figure}

During training we saved a checkpoint per epoch, and used the checkpoint with least validation loss for inference. The validation loss / epoch curves are given in Figure \ref{fig:validation_loss}.
During inference, we passed the Decision Context to the models and expected it to return the desired Design Decision.

\subsection{Metrics}
In accordance with NLP literature, the evaluation of text generation often relies on a combination of metrics rather than a single standard measure. In line with this practice, our evaluation incorporates ROUGE-1, BLEU score, METEOR, and BERTScore.

ROUGE (Recall-Oriented Understudy for Gisting Evaluation) \cite{b1} is a set of metrics used to evaluate the quality of machine-generated summaries. Here we are using ROUGE-1 that measures the overlap of unigrams (individual words) between the system-generated text and the reference text.

The BLEU (Bilingual Evaluation Understudy) \cite{b2} score is a metric used to evaluate the quality of machine-translated text.

METEOR (Metric for Evaluation of Translation with Explicit ORdering) \cite{b3} is a metric used for evaluating machine-generated text, particularly in the context of machine translation.

BERTScore \cite{b4} is an automatic evaluation metric used to assess the quality of text generation. It leverages pre-trained contextual embeddings from BERT (Bidirectional Encoder Representations from Transformers) \cite{b5} and measures the similarity between words in candidate and reference sentences using cosine similarity. BERTScore captures semantic similarity between texts has been shown to correlate well with human judgment in evaluating text generation outputs \cite{b4}. Hence we use it as the primary metric in this study.

Our experimental scripts and details can be found in:\\
\url{https://github.com/sa4s-serc/ArchAI_ADR}

\section{Results}\label{sec:results}

\begin{table*}[ht]
\centering
\caption{results 0-shot}
\begin{tabular}{c|ccc|ccc|c}
    \toprule
    \textbf{model} & \textbf{rouge-1} & \textbf{blue} & \textbf{Meteor} & & \textbf{BERTScore} & & support\\
    & & & & precision & recall & f1 &\\
    \midrule
    GPT2 & 0.071 & 0.008 & 0.109 & 0.740 & 0.808 & 0.772 & 78\\
    GPT2-medium & 0.071 & 0.007 & 0.103 & 0.734 & 0.81 & 0.769 & 78\\
    GPT2-large & 0.081 & 0.009 & 0.115 & 0.737 & 0.815 & 0.773 & 78\\
    GPT2-xl & 0.084 & 0.01 & 0.114 & 0.749 & 0.822 & 0.783 & 78\\
    GPT3-ada & 0.129 & 0.005 & 0.137 & 0.79 & 0.823 & 0.805 & 89\\
    GPT3-davinci & 0.154 & 0.009 & 0.167 & 0.799 & 0.837 & 0.817 & 89\\
    GPT3.5-text-davinci-003 & 0.242 & \textbf{0.031} & 0.198 & 0.846 & 0.849 & 0.847 & 91\\
    GPT-3.5-turbo & 0.235 & 0.029 & 0.213 & 0.836 & 0.848 & 0.841 & 92\\
    GPT-4 & \textbf{0.259} & 0.028 & \textbf{0.219} & 0.847 & \textbf{0.851} & \textbf{0.849} & 95\\
    \midrule
    T5-small & 0.203 & 0.017 & 0.168 & 0.84 & 0.831 & 0.835 & 91\\
    T5-base & 0.164 & 0.011 & 0.142 & 0.829 & 0.823 & 0.825 & 91\\
    T5-large & 0.16 & 0.014 & 0.137 & 0.816 & 0.823 & 0.819 & 91\\
    T5-3b & 0.088 & 0.004 & 0.077 & 0.794 & 0.81 & 0.801 & 91\\
    t0-3b & 0.187 & 0.005 & 0.122 & \textbf{0.856} & 0.823 & 0.839 & 91\\
    Flan-T5-small & 0.142 & 0.008 & 0.101 & 0.823 & 0.813 & 0.817 & 91\\
    Flan-T5-base & 0.158 & 0.012 & 0.115 & 0.845 & 0.822 & 0.833 & 91\\
    Flan-T5-large & 0.166 & 0.014 & 0.105 & 0.836 & 0.822 & 0.827 & 91\\
    Flan-T5-xl & 0.147 & 0.011 & 0.089 & 0.837 & 0.816 & 0.825 & 91\\
    \bottomrule
\end{tabular}
\label{tab:0shot}
\end{table*}

\begin{table*}[ht]
\centering
\caption{results few-shot}
\begin{tabular}{c|ccc|ccc|c}
    \toprule
    \textbf{model} & \textbf{rouge-1} & \textbf{blue} & \textbf{Meteor} & & \textbf{BERTScore} & & support\\
    & & & & precision & recall & f1 &\\
    \midrule
    GPT2 & 0.087 & 0.012 & 0.135 & 0.778 & 0.841 & 0.807 & 13\\
    GPT2-medium & 0.11 & 0.016 & 0.017 & 0.785 & 0.837 & 0.809 & 13\\
    GPT2-large & 0.117 & 0.018 & 0.176 & 0.79 & 0.851 & 0.818 & 13\\
    GPT2-xl & 0.122 & 0.018 & 0.18 & 0.794 & \textbf{0.857} & 0.823 & 13\\
    GPT3-ada & 0.14 & 0.008 & 0.158 & 0.789 & 0.827 & 0.807 & 86\\
    GPT3-davinci & 0.171 & 0.011 & 0.181 & 0.802 & 0.84 & 0.82 & 86\\
    GPT3.5-text-davinci-003 & \textbf{0.245} & 0.028 & 0.207 & 0.\textbf{849} & 0.851 & \textbf{0.849} & 91\\
    GPT-3.5-turbo & 0.226 & 0.027 & \textbf{0.219} & 0.832 & 0.85 & 0.84 & 92\\
    GPT-4 & 0.23 & \textbf{0.03} & 0.21 & 0.836 & 0.851 & 0.843 & 93\\
    \midrule
    T5-small & 0.151 & 0.007 & 0.121 & 0.824 & 0.819 & 0.821 & 91\\
    T5-base & 0.146 & 0.008 & 0.12 & 0.805 & 0.821 & 0.812 & 91\\
    T5-large & 0.177 & 0.014 & 0.146 & 0.811 & 0.825 & 0.817 & 91\\
    T5-3b & 0.136 & 0.006 & 0.1 & 0.805 & 0.821 & 0.812 & 91\\
    t0-3b & 0.152 & 0.011 & 0.133 & 0.808 & 0.82 & 0.813 & 91\\
    Flan-T5-small & 0.162 & 0.005 & 0.152 & 0.782 & 0.824 & 0.801 & 91\\
    Flan-T5-base & 0.176 & 0.011 & 0.158 & 0.811 & 0.829 & 0.819 & 91\\
    Flan-T5-large & 0.168 & 0.014 & 0.16 & 0.81 & 0.828 & 0.818 & 91\\
    Flan-T5-xl & 0.17 & 0.011 & 0.155 & 0.805 & 0.825 & 0.814 & 91\\
    \bottomrule
\end{tabular}
\label{tab:few_shot}
\end{table*}

\begin{table*}[ht]
\centering
\caption{results fine-tuning}
\begin{tabular}{c|ccc|ccc|cc}
    \toprule
    \textbf{model} & \textbf{rouge-1} & \textbf{blue} & \textbf{Meteor} & & \textbf{BERTScore} & & & support\\
    & & & & precision & recall & f1 & Train & Test\\
    \midrule
    GPT2 & 0.155 & 0.01 & 0.176 & 0.797 & 0.834 & 0.815 & 62 & 16\\
    T5-small & 0.199 & 0.008 & 0.156 & 0.835 & 0.824 & 0.829 & 73 & 19\\
    T5-base & 0.195 & 0.022 & 0.168 & 0.839 & 0.838 & 0.838 & 71 & 18\\
    Flan-T5-small & 0.19 & 0.018 & 0.139 & 0.842 & 0.822 & 0.831 & 73 & 19\\
    Flan-T5-base & \textbf{0.231} & \textbf{0.028} & \textbf{0.171} & \textbf{0.842} & \textbf{0.841} & \textbf{0.841} & 71 & 18\\
    \bottomrule
\end{tabular}
\label{tab:finetune}
\end{table*}

\begin{figure*}[t]
    \centering
    \includegraphics[width=\textwidth]{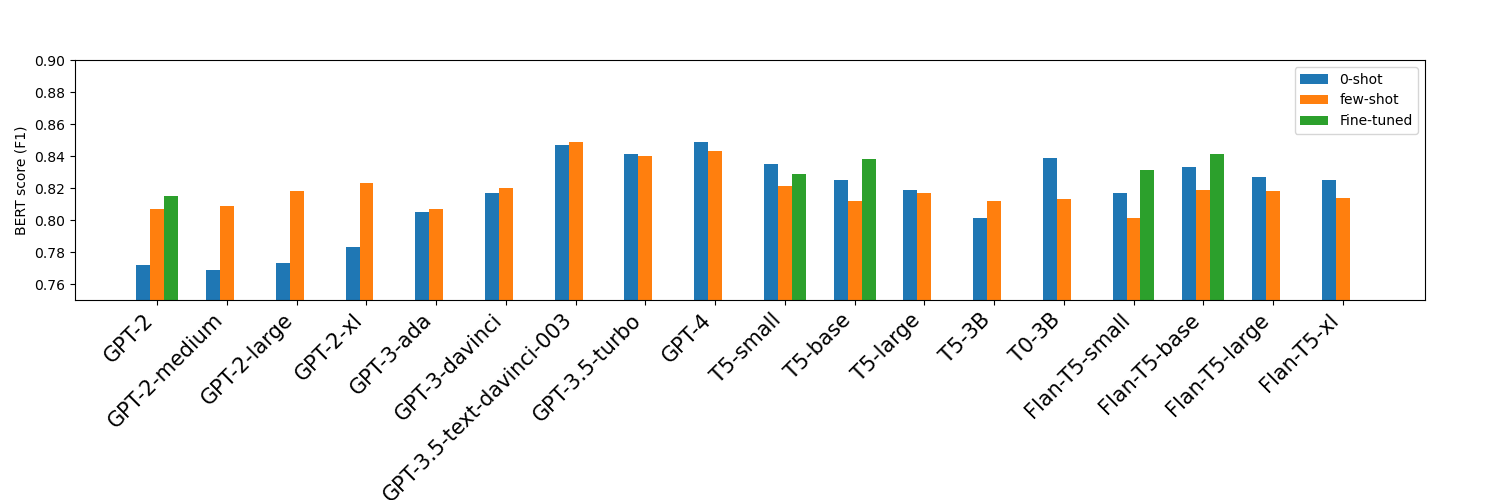}
    \caption{BERTScore (F1)}
    \label{fig:BERT_score_F1}
\end{figure*}

In this section, we report the results of our empirical experimentation according to the research questions guiding this study (see Section \ref{sec:study_design}).

Tables~\ref{tab:0shot}, \ref{tab:few_shot} and \ref{tab:finetune} present the metric values for the 0-shot, few-shot, and fine-tuning experiments, respectively, with the highest value for each metric emphasized in bold. The BERTScore (F1), which is our primary metric, is graphically represented in Figure \ref{fig:BERT_score_F1}. Although there are instances where the metrics do not align, the result tables show that they generally work in tandem and show a correlation with the BERTScore (F1).

In addition to computing evaluation metrics, we manually reviewed a subset of the generated samples, focusing on both content and formatting of the generated Decisions. Sample generated Decision by all three approaches are given in Figure \ref{fig:Sample_Generated_Decision}, following the same sample from Figure 2.

\begin{figure}[t]
\centering
\fbox{
    \begin{minipage}{0.45\textwidth}
        \textbf{Actual Decision}\\
        We have decided to use Python as our primary programming language for our project.
        \\
        
        \textbf{Decision Generated by GPT-4 in 0-shot}\\
        We will use Python as our programming language for this project. Python has emerged as a reliable, versatile language with a rich ecosystem of libraries and frameworks that cater to our specific needs including data analysis (Pandas, NumPy), machine learning (scikit-learn, TensorFlow), and web development (Django, Flask). Its simplicity and readability make for quicker turnaround times in development phases. Therefore, Python fits perfectly with the requirements of our project, providing us with all the tools necessary for successful development and deployment.
        \\

        \textbf{Decision Generated by text-davinci-003 in few-shot}\\
        After considering the various criteria discussed, we have decided to go ahead and use Python for our project.

        Python is a versatile, powerful language with a wide range of features, libraries, and tools that make it ideal for multiple tasks, including data analysis, machine learning, and web development. It is simple to learn and use, and there is a large community of developers who can provide support and resources. Python is also open-source, free to use, and has a great number of packages available for various tasks.
        
        Python is also a great choice in terms of performance and scalability. With the right packages, it can be used for big data processing and high performance applications. Additionally, Python is an interpreted language, which makes it easier to debug and maintain than compiled languages.
        
        Overall, Python is a great choice for our project, given its features and capabilities, scalability, and community support.
        \\
        
        \textbf{Decision Generated by Flan-T5-small after fine-tuning}\\
        It will have a large dataset, which will display a lot of data. This can help us in preparing the application. We need to choose Python as a programming language, because all of our projects will require Python to use it.
        
    \end{minipage}
}
\caption{Sample Generated Decision}
\label{fig:Sample_Generated_Decision}
\end{figure}

\subsection{Results $RQ_1$: Can LLMs be successfully employed to generate architectural design decisions from a given context in a zero-shot setting?}

In this introductory research question, we determine  whether an architect can effectively employ an LLM by providing it with a decision Context to derive the corresponding Design Decision.

While experimenting with various prompts to find the most effective prompt, we observed including the term 'Architectural Decision Record' in the prompt enhances the overall performance in 0-shot prompting for all models.
However, this improvement pertains for adhering to the ADR markdown format rather than impacting the content of the Design Decision the model is required to produce. We also find including the \textit{Title} (e.g., 'MySQL database') of the ADR does not yield notable improvements in the results for 0-shot prompting across all the models. We've also noticed that larger and more descriptive contexts lead to better and more comprehensive decisions.

From Table~\ref{tab:0shot} and Figure \ref{fig:BERT_score_F1} it is evident that the large GPT models and T0 performs well in 0-shot approach, while the T5 models show mediocre results, and the smaller GPT models performs the worst. GPT-4 scores 0.849 as BERTScore (F1) stands out to be the best model with best performance in all metrics other than Blue score and BERTScore (precision). 

Upon manually inspecting the generated content, it becomes evident that larger GPT models consistently produce pertinent responses in the correct ADR format, with GPT-4's formatting approaching human level. T0 also generates relevant responses but lacks proper formatting. Conversely, the remaining models demonstrate suboptimal performance in both generating accurate decisions and formatting them correctly, with smaller GPT models frequently producing hallucinatory content.

\subsection{Results $RQ_2$: Does few-shot approach affect or improve a LLM’s ability to generate Design Decisions?}

In $RQ_2$ we aim to assess the capability of LLMs in generating Design Decisons in few-shot approach. Our goal is to access if smaller, more economical LLMs can outperform larger models in zero-shot scenarios by integrating sample Context-Decision pairs into the prompts.

Based on the data presented in Table~\ref{tab:few_shot} and the observations from Figure \ref{fig:BERT_score_F1}, it is apparent that the larger GPT models excel when employing a few-shot approach, whereas the performance of other models remains moderate. Notably, GPT-3.5's text-davinci-003 stands out as the top performer in this approach, attaining the highest scores in rouge-1, BERTScore (precision), and BERTScore (F1). Specifically, it achieves a BERTScore (F1) of 0.849, which matches the performance of GPT-4 in the 0-shot setting.

Upon manually examining the generated content, we observe that the quality of generation closely resembles that of the 0-shot approach, wherein larger GPT models consistently produce relevant Design Decisions with appropriate formatting, in contrast to other models that exhibit shortcomings in this regard. Notably, smaller GPT models stand out as an exception. Although these smaller models frequently generated hallucinatory content in the 0-shot approach, there is a significant reduction in hallucinations when employing the few-shot approach. The rise in metric values, as illustrated in Figure \ref{fig:BERT_score_F1}, further attests to this observation.

\subsection{Results $RQ_3$: Does Fine-tuning LLMs enhances it’s capability of generating architectural Design Decisions based on a provided context?}

$RQ_3$ examines whether fine-tuning LLMs with Context-Decision pairs improves their ability to generate Design Decisions. We explore if smaller locally trainable and deployable fine-tuned models can match the capability of extensive cloud-based models in generating comparable Design Decisions.

Table~\ref{tab:finetune} and Figure \ref{fig:BERT_score_F1} illustrate that all the fine-tuned models outperform  their off-the-shelf counterparts, with T5 being the only exception. Flan-T5 emerges as the leading fine-tuned model across all metrics, achieving an impressive score of 0.841 for BERTScore (F1).

On manual examination of the generated content, we find that the T5 models stop hallucinating after fine-tuning. Additionally, the decision content also improves after fine-tuning.

\section{Discussion}\label{sec:discussion}
This exploratory study provides empirical evidence that LLM's can be leveraged to Generate Architecture Design Decision given the Context, and would definitely help Software Architects in generating ADR's. Below, we address each research question by examining the outcomes of our experiments. Several conclusions are drawn regarding the efficiency of using LLMs to generate ADRs.

\subsection{Can LLMs be successfully employed to generate architectural design decisions from a given context in a zero-shot setting? ($RQ_1$)}

Yes, LLMs can be used to generate Design Decisions in zero-shot setting, but not entirely autonomously. Even the best performing models fails to generate the required Decision comprehensively, necessitating human involvement.

Notably, smaller models such as GPT-2, T5, and Flan-T5, originally not designed for zero-shot capabilities, exhibit subpar performance. T0, despite being a relatively smaller model, returns relevant decisions but lacks adherence to the ADR format. Contrastingly, smaller GPT models performs better in adhering to correct ADR format than producing correct response. GPT-4 correctly adheres to ADR format but falls short in generating  the required Decision comprehensively. Both manual observation and evaluation metrics support these findings, necessitating further research on leveraging LLMs to reach human-level proficiency in generating Design Decisions.

It's important to highlight that both GPT-3.5 and GPT-4 are cloud-based and offered as 'Software as a Service.' Consequently, to use these models, data must be transmitted to the service provider, like OpenAI\footnote{https://platform.openai.com/docs/models}, potentially raising concerns related to privacy, intellectual security, or legal implications.

Hence we recommend that if security is not a concern, LLMs like GPT-3.5 and GPT-4 can be successfully employed to generate Design Decisions, but not autonomously. Rather it should be used to assist Architects in documenting and making Design Decisions.

\begin{figure}[h]
\centering
\fcolorbox{black}{ash}{
    \begin{minipage}{0.45\textwidth}
        \textbf{Main findings $RQ_1$ (Generating Design Decisions in 0-shot approach):} While LLMs can generate Design Decisions using a 0-shot approach, the quality does not reach human-level. Nonetheless, this method can be employed to support architects in documenting and formulating Design Decisions.
    \end{minipage}
}
\end{figure}

\subsection{Does few-shot approach affect or improve a LLM’s ability to generate Design Decisions? ($RQ_2$)}

The impact of the few-shot approach compared to the 0-shot approach is inconclusive. As depicted in Figure 9, while the few-shot approach enhances performance for GPT-2 models, in certain models such as GPT-3, GPT-3.5 and GPT-4 the effect is insignificant. Interestingly, for some models like the entire T5 series (excluding T5-3B), the performance actually decreases. This observation is consistent across other metrics. Manual observations suggests that neither adherence to the ADR format nor the content of the generated Decision improves significantly with the few-shot approach for any of the models. The only exception is GPT-2, where the few-shot approach reduces hallucinations.

Based on the metrics, it can be inferred that the performance of GPT-3.5's text-davinci-003 using the few-shot approach is equal to that of GPT-4 using the 0-shot approach, both of which exhibit the best metrics in our study. However, it's important to note that GPT-4 is significantly larger (10 times) and expensive (1.5 times) compared text-davinci-003.

Hence we infer from this study that one may use smaller and cheaper models like GPT-3.5's text-davinci-003 in a few shot approach and generate Design Decisions of quality at par with powerful models like GPT-4 at a cheaper rate.

\begin{figure}[h]
\centering
\fcolorbox{black}{ash}{
    \begin{minipage}{0.45\textwidth}
        \textbf{Main findings $RQ_2$ (Impact of few-shot approach in Generating Design Decisions):} While the performance of certain LLMs may show improvement in a few-shot setting, the overall phenomenon lacks generalization and remains inconclusive. Nevertheless, smaller LLMs, when presented with a few-shot prompt, can be employed as substitutes for larger LLMs in certain scenarios.
    \end{minipage}
}
\end{figure}

\subsection{Does Fine-tuning LLMs enhances it’s capability of generating architectural Design Decisions based on a provided context? ($RQ_3$)}

It is conclusive from the metrics that fine-tuning does improve the Design Decisions generating capability of LLMs, with T5-small being the only exception. Manual observation also suggests the same both with respect to formatting and content. Though it's observed that even the best performing fine-tuned model can't match the performance of GPT-4 in terms of following ADR markdown format.

The metrics show that after fine-tuning, the top-performing model, Flan-T5-base, with 248 million parameters, achieves results comparable with GPT-3.5 in a few-shot approach, which has 175 billion parameters (700 times more). Additionally, fine-tuned Flan-T5-base also produces results comparable to GPT-4 in a 0-shot approach, boasting over a trillion parameters (7000 times more). This clearly demonstrates that smaller fine-tuned models can effectively substitute larger models for generating design decisions. Due to hardware limitations, our fine-tuning efforts were capped at models with fewer than 400 million parameters. The results suggest that T5-3B or Flan-T5-xl, each with 3 billion parameters (30 times less than GPT-3.5), could potentially match GPT-4 performance through fine-tuning. This indicates that a more extensive study with increased data and hardware capabilities is needed which may yield more promising outcomes.

It is crucial to emphasize that smaller models, such as Flan-T5, can be hosted in-house in most corporate environments, thereby mitigating potential privacy, security, or legal concerns.

Hence, with respect to $RQ_3$ we conclude, while smaller fine-tuned models like Flan-T5 may not reach the performance level of giants like GPT-4 in generating design decisions, they serve as valuable alternatives in scenarios where privacy or security considerations are paramount.

\begin{figure}[h]
\centering
\fcolorbox{black}{ash}{
    \begin{minipage}{0.45\textwidth}
        \textbf{Main findings $RQ_3$ (Impact of Fine-tuning in Generating Design Decisions):} Fine-tuned LLMs exhibit improved capability in generating design decisions. Compact fine-tuned models, which require minimal infrastructure for hosting, demonstrate results comparable with extensive LLMs and can be utilized as their substitutes in specific scenarios.
    \end{minipage}
}
\end{figure}

\section{Threats to validity}\label{sec:threats_to_validity}
\noindent In this section, we discuss threats to the validity of our study, following the categorization provided by Wholin et al. \cite{b7}.

\subsection{Internal Validity}
A threat to internal validity may arise from the selection of metrics, given that evaluating text generation is a challenging and unresolved problem. To mitigate this threat, we have addressed it by adopting a set of metrics commonly utilized by the NLP community. These metrics aim to capture the quality of generated text from various perspectives.

An additional threat to internal validity could arise from inconsistent ADR writing styles, potentially leading to inaccurate evaluations. Different organizations or individuals may employ varied styles, such as precise or descriptive approaches. Even if the LLM generates a matching Design Decision, evaluation metrics may suffer if the wording or style doesn't align. To mitigate this, we've introduced diverse metrics capturing text quality from various perspectives. Additionally, we've incorporated BERTScore, evaluating semantic meaning rather than word-to-word comparison, to mitigate stylistic impacts.

\subsection{Construct Validity}
A threat to construct validity is the limited amount of data used for training while addressing $RQ_3$. This threat remains partially mitigated due to unavailability of open source ADR repositories on the internet. Despite this challenge, we have gathered 95 ADRs from 5 different repositories.
While fine-tuning, we used 80\% of the data for training, and rest 20\% for testing. It should be mentioned that we used the same data for validation and testing for the fine-tuning experiments as explained in Section \ref{sec:study_design} due to the low amount of data available. This can be considered a limitation to this study, though it's unlikely that it could change the conclusions.

\subsection{External Validity}
A potential threat to our study's external validity stems from LLM selection. Due to the multitude of available LLMs, it's impractical to include all. To address this, we carefully select representative LLMs, incorporating diverse models in terms of availability, size, training data, techniques, and interaction styles. Further details is given in Section~\ref{sec:study_design}.

An additional concern regarding External Validity is the potential that the gathered ADRs for this study may not be  representative of ADRs in a broader context. To address this concern, we conducted 
web searches on ADR and selected 5 repositories with well documented ADRs. Further details is given in Section~\ref{sec:study_design}

\section{Conclusion and Future work}\label{sec:conclusion}
This study explores the potential of leveraging LLMs, specifically GPT and T5-based models, to automate the generation of ADRs, which are a crucial part of AKM. The investigation involves using 0-shot, few-shot, and fine-tuning approaches to generate ADR Decisions based on their respective Contexts.
The study reveals that LLMs demonstrate the capability to generate ADD. Models like GPT-3.5, GPT-4, T0 effectively generate relevant Design Decisions adhering to ADR markdown format with both 0-shot and few-shot prompting. Fine-tuning enhances the performance of all models significantly. GPT-4 excels in 0-shot prompting, while even smaller models like text-davinci-003 yield similar results using a few-shot approach. Smaller models such as Flan-T5-base demonstrate comparable results after fine-tuning.
In summary, LLMs may not be entirely dependable for ADR generation, but they can effectively assist architects in documenting and formulating Design Decisions. Smaller fine-tuned models, requiring minimal hosting, can be locally employed for Decision generation, especially in privacy or security-sensitive scenarios.

Future work involves fine-tuning large LLMs to enhance Design Decision generation, aiming for human-level performance. 
In most of the experiments we found LLMs are not able to comprehensively capture the Design Decisions as per human level. This may improve it the LLM had access to more world data or knowledge such as other ADRs. Retrieval Augmented Generation might help here.
Additionally, in vivo experiments with architects in live projects will assess the effectiveness of auto-generating ADRs for widespread adoption. Recognizing challenges in fully capturing Design Decisions from decision Context, future efforts will focus on integrating contextual information from diverse sources, including other ADRs, AKM components like design diagrams, and the codebase, to improve the generation process.


\begin{thebibliography}{00}

\bibitem{b13} Tang, Antony \& Avgeriou, Paris \& Jansen, Anton \& Capilla, Rafael \& Ali Babar, Muhammad. (2013). A comparative study of architecture knowledge management tools. Journal of Systems and Software. 352-370. 10.1016/j.jss.2009.08.032.

\bibitem{b15} Weinreich, Rainer \& Groher, Iris. (2016). Software architecture knowledge management approaches and their support for knowledge management activities: A systematic literature review. Information and Software Technology. 80. 10.1016/j.infsof.2016.09.007.

\bibitem{b14} Capilla, Rafael \& Jansen, Anton \& Tang, Antony \& Avgeriou, Paris \& Ali Babar, Muhammad. (2016). 10 years of Software Architecture Knowledge Management: Practice and Future. Journal of Systems and Software. 10.1016/j.jss.2015.08.054.

\bibitem{b12} Jansen, Anton \& Bosch, Jan. (2005). Software Architecture as a Set of Architectural Design Decisions. Proceedings - 5th Working IEEE/IFIP Conference on Software Architecture, WICSA 2005. 2005. 109-120. 10.1109/WICSA.2005.61.

\bibitem{b27} Buchgeher, Georg \& Schöberl, Stefan \& Geist, Verena \& Dorninger, Bernhard \& Haindl, Philipp \& Weinreich, Rainer. (2023). Using Architecture Decision Records in Open Source Projects – An MSR Study on GitHub. IEEE Access. 11. 63725 - 63740. 10.1109/ACCESS.2023.3287654.

\bibitem{b29} Marius Mosbach, Tiago Pimentel, Shauli Ravfogel, Dietrich Klakow, and Yanai Elazar. 2023. Few-shot Fine-tuning vs. In-context Learning: A Fair Comparison and Evaluation. In Findings of the Association for Computational Linguistics: ACL 2023, pages 12284–12314, Toronto, Canada. Association for Computational Linguistics.

\bibitem{b8} Ashish Vaswani, Noam Shazeer, Niki Parmar, Jakob Uszkoreit, Llion Jones, Aidan N. Gomez, Łukasz Kaiser, and Illia Polosukhin. 2017. Attention is all you need. In Proceedings of the 31st International Conference on Neural Information Processing Systems (NIPS'17). Curran Associates Inc., Red Hook, NY, USA, 6000–6010.

\bibitem{b5} Devlin, Jacob \& Chang, Ming-Wei \& Lee, Kenton \& Toutanova, Kristina. (2018). BERT: Pre-training of Deep Bidirectional Transformers for Language Understanding.

\bibitem{b10} Raffel, Colin, Noam M. Shazeer, Adam Roberts, Katherine Lee, Sharan Narang, Michael Matena, Yanqi Zhou, Wei Li and Peter J. Liu. “Exploring the Limits of Transfer Learning with a Unified Text-to-Text Transformer.” J. Mach. Learn. Res. 21 (2019): 140:1-140:67.

\bibitem{b11} Radford, Alec \& Narasimhan, Karthik \& Salimans, Tim \& Sutskever, Ilya. "Improving language understanding by generative pre-training." (2018):

\bibitem{b28} Brown, Tom \& Mann, Benjamin \& Ryder, Nick \& Subbiah, Melanie \& Kaplan, Jared \& Dhariwal, Prafulla \& Neelakantan, Arvind \& Shyam, Pranav \& Sastry, Girish \& Askell, Amanda \& Agarwal, Sandhini \& Herbert-Voss, Ariel \& Krueger, Gretchen \& Henighan, Tom \& Child, Rewon \& Ramesh, Aditya \& Ziegler, Daniel \& Wu, Jeffrey \& Winter, Clemens \& Amodei, Dario. (2020). Language Models are Few-Shot Learners.

\bibitem{b23} Shahin, Mojtaba \& Liang, Peng \& Khayyambashi, Mohammad. (2009). Architectural Design Decision: Existing Models and Tools. 293-296. 10.1109/WICSA.2009.5290823.

\bibitem{b24} Scheerer, Max \& Busch, Axel \& Koziolek, Anne. (2017). Automatic evaluation of complex design decisions in component-based software architectures. 67-76. 10.1145/3127041.3127059. 

\bibitem{b25} Shahbazian, Arman \& Lee, Youn \& Le, Duc \& Brun, Yuriy \& Medvidovic, Nenad. (2018). Recovering Architectural Design Decisions. 95-9509. 10.1109/ICSA.2018.00019.

\bibitem{b16} Shafiq, Saad \& Mashkoor, Atif \& Dorn, Christoph \& Egyed, Alexander. (2020). Machine Learning for Software Engineering: A Systematic Mapping.

\bibitem{b17} Meinke, Karl \& Bennaceur, Amel. (2017). Machine Learning for Software Engineering Models, Methods, and Applications. 10.1145/3183440.3183461.

\bibitem{b18} Wang, April \& Wang, Dakuo \& Drozdal, Jaimie \& Muller, Michael \& Park, Soya \& Weisz, Justin \& Liu, Xuye \& Wu, Lingfei \& Dugan, Casey. (2022). Documentation Matters: Human-Centered AI System to Assist Data Science Code Documentation in Computational Notebooks. ACM Transactions on Computer-Human Interaction. 29. 1-33. 10.1145/3489465. 

\bibitem{b21} Gu, Jian \& Salza, Pasquale \& Gall, Harald. (2022). Assemble Foundation Models for Automatic Code Summarization. 935-946. 10.1109/SANER53432.2022.00112.

\bibitem{b19} Mu, Fangwen \& Chen, Xiao \& Shi, Lin \& Wang, Song \& Wang, Qing. (2023). Developer-Intent Driven Code Comment Generation.

\bibitem{b20} Li, Xueying \& Liang, Peng \& Li, Zengyang. (2020). Automatic Identification of Decisions from the Hibernate Developer Mailing List. 10.1145/3383219.3383225.

\bibitem{b26} Bhat, Manoj \& Tinnes, Christof \& Shumaiev, Klym \& Biesdorf, Andreas \& Hohenstein, Uwe \& Matthes, Florian. (2019). ADeX: A Tool for Automatic Curation of Design Decision Knowledge for Architectural Decision Recommendations. 10.1109/ICSA-C.2019.00035.

\bibitem{b6} V. R. Basili, G. Caldiera, and D. Rombach, “The Goal Question Metric
Approach,” in Encyclopedia of Software Engineering. Wiley, 1994, pp.
528–532.

\bibitem{b1} Chin-Yew Lin. 2004. ROUGE: A Package for Automatic Evaluation of Summaries. In Text Summarization Branches Out, pages 74–81, Barcelona, Spain. Association for Computational Linguistics.

\bibitem{b2} Papineni, Kishore \& Roukos, Salim \& Ward, Todd \& Zhu, Wei Jing. (2002). BLEU: a Method for Automatic Evaluation of Machine Translation. 10.3115/1073083.1073135.

\bibitem{b3} Lavie, Alon \& Agarwal, Abhaya. (2007). METEOR: An automatic metric for MT evaluation with high levels of correlation with human judgments. 228-231. 

\bibitem{b4} Zhang, Tianyi \& Kishore, Varsha \& Wu, Felix \& Weinberger, Kilian \& Artzi, Yoav. (2019). BERTScore: Evaluating Text Generation with BERT.

\bibitem{b7} C. Wohlin, P. Runeson, M. Host, M. C. Ohlsson, B. Regnell, and A. Wesslen, Experimentation in software engineering. Springer Science \& Business Media, 2012

\end{thebibliography}
\end{document}